\documentclass[preprint,12pt]{elsarticle}
\usepackage{amsfonts,amssymb}
\usepackage[fleqn]{amsmath}
\def \sax {{\it Beppo}SAX}
\def \grss {GRS 1915+105~}
\def \grs {GRS 1915+105}

\journal{International Journal of Non-Linear Mechanics}
\begin{document}
\begin{frontmatter}
\title {Lyapunov functions for a non-linear model of the X-ray bursting of the microquasar GRS 1915+105}

\author{A. Ardito} 

\address{Department of Mathematics G. Castelnuovo, Sapienza Roma University, Roma, Italy}

\author{P. Ricciardi}

\address{Department of Mathematics G. Castelnuovo, Sapienza Roma University, Roma, Italy \\
         In Unam Sapientiam, Roma, Italy }

\author{E. Massaro}

\address{IAPS, Istituto Nazionale di Astrofisica INAF, Roma, Italy \\
         In Unam Sapientiam, Roma, Italy }

\author{T. Mineo}

\address{IASF, Istituto Nazionale di Astrofisica INAF, Palermo, Italy  }

\author{F. Massa}

\address{ (retired) INFN, Sezione Roma1, Roma, Italy }

\begin{abstract}
This paper introduces a biparametric family of Lyapunov functions for a non-linear 
mathematical model based on the FitzHugh-Nagumo equations able to reproduce some 
main features of the X-ray bursting behaviour exhibited by the microquasar \grs.
These functions are useful to investigate the properties of equilibrium points and 
allow us to demonstrate a theorem on the global stability. 
The transition between bursting and stable behaviour is also analyzed. 

\end{abstract}

\begin{keyword}
Non-linear dynamic stability; Lyapunov function; Microquasar oscillations
\end {keyword}

\end{frontmatter}

\section{\label{sec_1}Introduction}
The X-ray source \grs, the prototype of microquasars, was discovered by 
Castro-Tirado et al. [1], and later identified with a binary system containing 
an accreting disk around a black hole by Mirabel and Rodr\'iguez [2].
The X-ray emission is highly variable and characterised by a several different 
variability patterns, that alternate steady and noisy emission to series of 
bursts and dips.  
A first classification of the observed time behaviour based on the signal 
structure and on the photon energy distribution was presented by Belloni et al. 
[3], who defined twelve classes identified by a greek letter, but more patterns 
were observed on subsequent occasions. 
In particular, the variability class $\rho$ is characterised by series of bursts 
with a variable recurrence time and superposed to a rather stable level. 
The time profile of the bursts presents a rather smooth rising branch, the {\it Slow 
Leading Trail} (hereafter SLT), followed by a {\it Pulse} with a few intense and short 
peaks and a fast decline, as those shown in the upper panel of Fig. 1.
The burst sequence was soon recognized as an evidence of a limit-cycle by Taam et 
al. [4] and explained by the occurrence in an physical parameters' space, like the 
plane of the disk temperature $vs$ the integrated surface density, of an 
instability region  where closed trajectories can be established (see, for instance, 
Taam \& Lin [5], Lasota \& Pelat [6], Mineshige [7], Szuszkiewicz \& Miller [8]).
However, the application of these models to observational results requires numerical 
solution of a complex system of non-linear partial differential equations taking 
into account several non-observable quantities in the viscous fluid dynamics and 
thermal processes in the accreting gas orbiting around a black hole.
This approach makes hard to establish a simpler mathematical modelling of 
the source's behaviour and the transition from a variability class to another.
We therefore tried a different way based on a direct description of the 
observational data by means of mathematical models.

\begin{figure}
\centering
\includegraphics[width=6.5cm,angle=0]{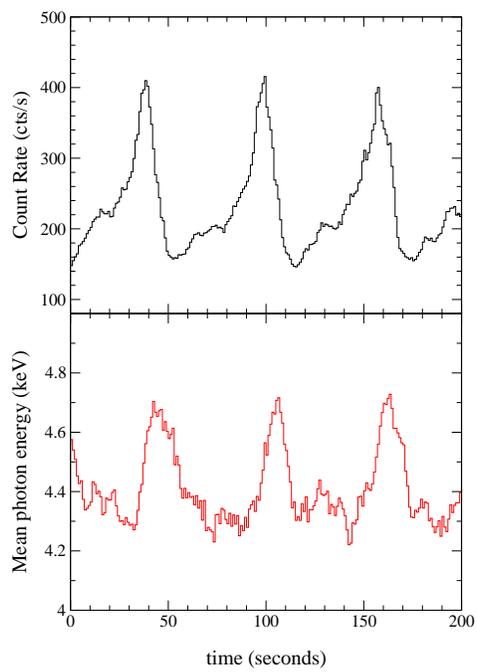}
\caption{\label{CHS_fig1} The top panel shows a short segment of the count rate 
time series in the entire MECS energy range during the October 2000 observation 
of \sax~ [9]; the bottom panel presents the simultaneous changes of mean energy.
}
\label{fig1}
\end{figure}

In a couple of papers, Massaro et al. [9] and Massa et al. [10] studied the time properties 
of a long series of bursts of \grss having a variable recurrence ranging time from 40 to 
more than 100 seconds. 
Massa et al. [10] developed an phase averaging algorithm to derive the burst mean profile 
and the mean photon energy curve which are shown in the left panel of Fig. 2. 
On this basis Massaro et al. [11] were able to describe this cycle by means of a modified 
FitzHugh-Nagumo (FitzHugh [12], Nagumo et al. [13]) model and found that it reproduces the 
mean burst shape and some other observational properties of the $\rho$ class.
In the same paper [11] we also demonstrated that a limit cycle can exist and gave the 
conditions for the parameters to have this solution.
It is given by the following system in the two dimensionless variables $x(t)$ and $y(t)$:

\begin{equation}
\left\{
\begin{array}{lcl}
dx/dt &=& - \rho x^3 + \chi x - \gamma y -  J  \\ 
dy/dt &=& x - y  
\end{array} \right.
\end{equation}

\noindent
where
\begin{eqnarray}
(i)  &~~~~ \rho , \chi , \gamma ~and~ J ~are~positive~parameters \nonumber \\
(ii) &~~~~ \chi < \gamma + 3 (\rho J^2/4)^{1/3} ~~,~~ \gamma > 1 
\end{eqnarray}

The parameter $J$ can be considered as an external forcing quantity, whereas $\rho$, 
$\chi$ and $\gamma$ can be related to the physical structure of the accreting disk.

\begin{figure} 
\includegraphics[height=6.9cm,angle=-90,scale=1.0]{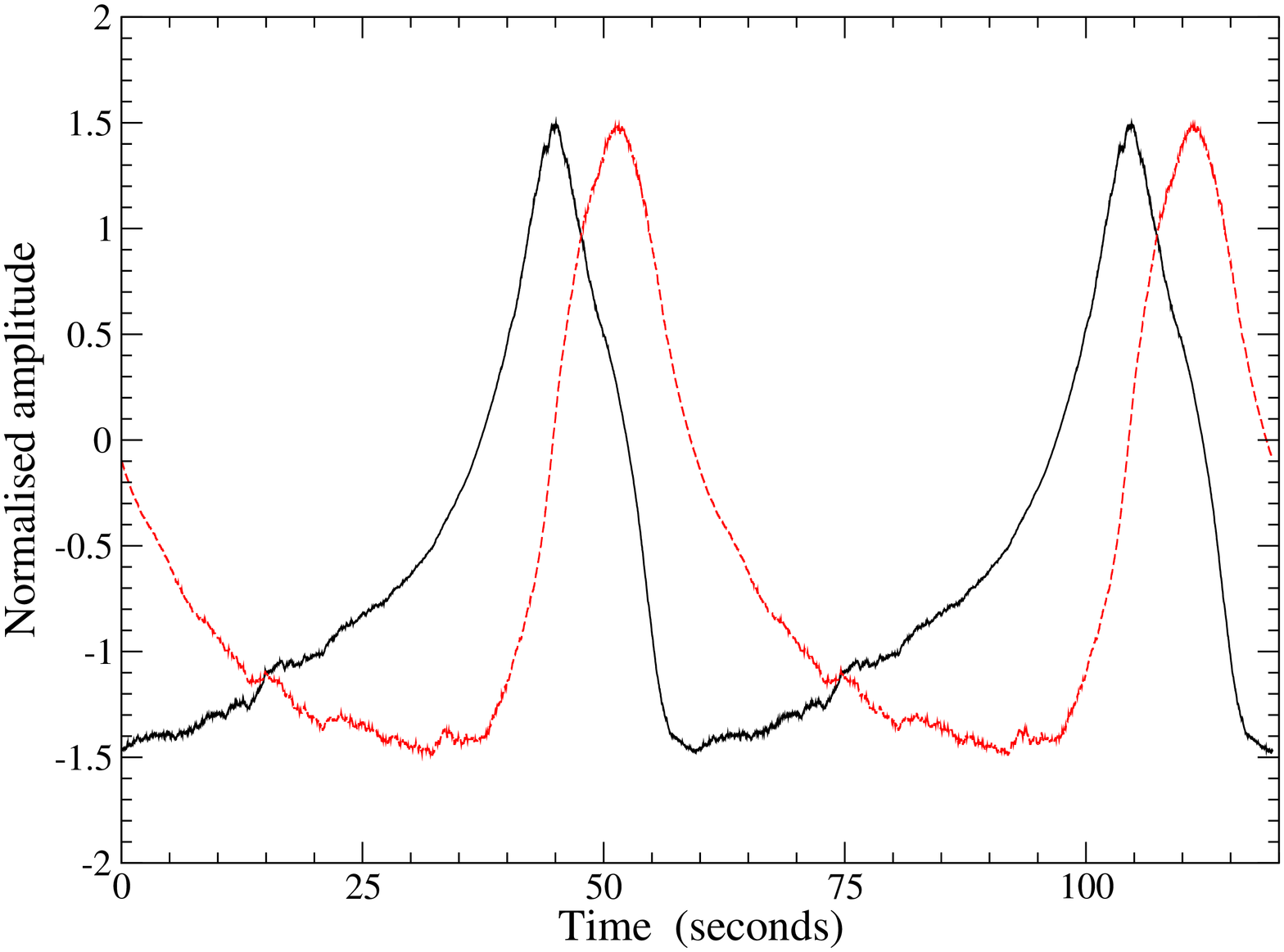}
\includegraphics[height=6.9cm,angle=-90,scale=1.0]{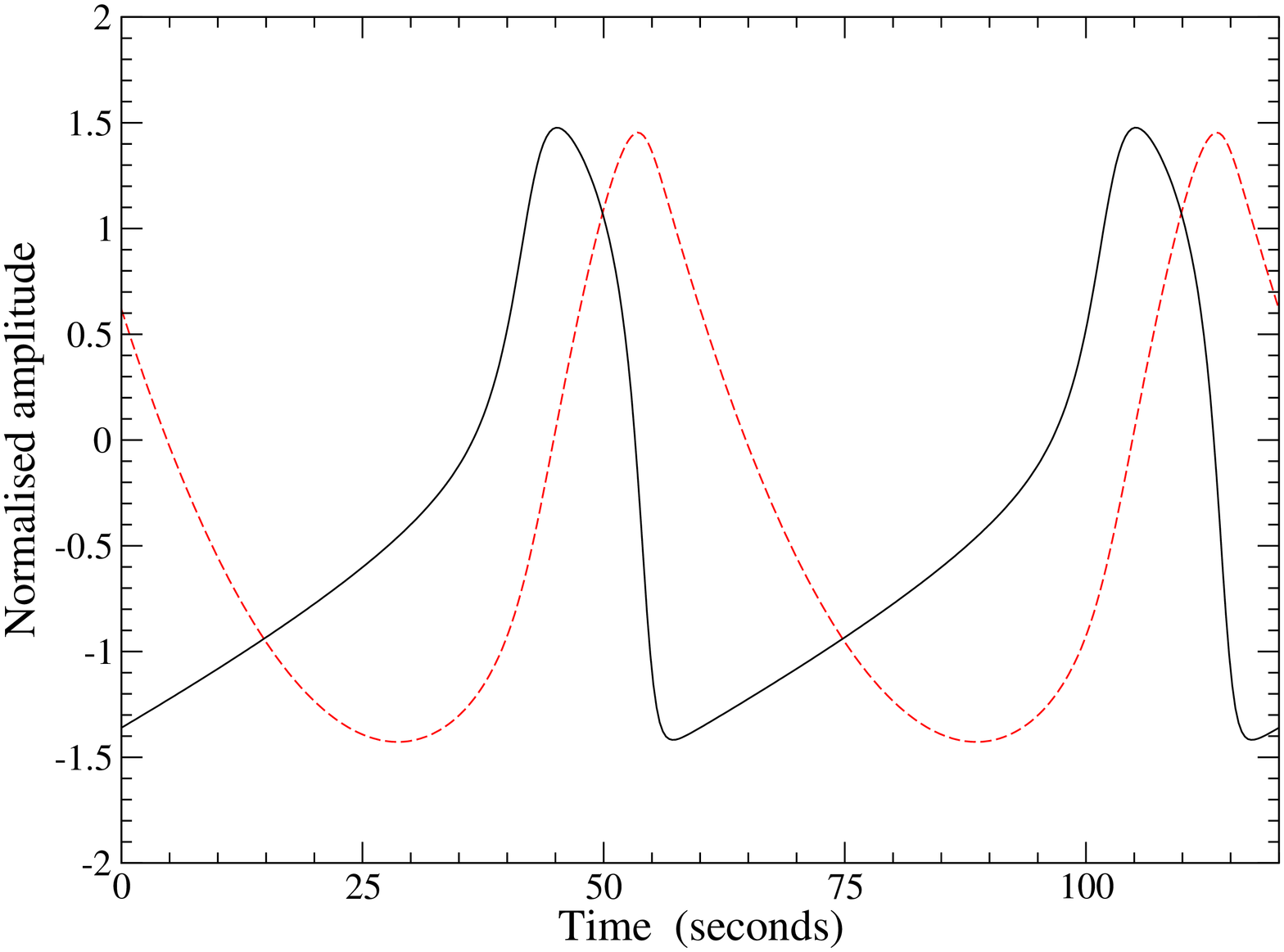}
\caption[]{
Left panel: time evolution of the mean pulse profile (black continuous line) and mean photon energy (red dashed line)
of \grss in the $\rho$ class.~
Right panel: results obtained from numerical solution of the model of Eq. (1) with 
the parameters' value given in the text, black continuous and red dashed lines represent the 
time behaviour of the $x$ and $y$ variables, respectively scaled as the curves in the 
left panel.
}
\label{fig2}
\end{figure}

The success of ~the system (1)~ in reproducing ~the \grss 

\noindent
burst series depends on the presence of the cubic term that could 
be related to radiative energy dissipation of the accretion disk considering,
for instance, that the black body photon emissivity is also proportional to the third 
power of the temperature.
These equations, however, are not the unique for this goal:
for instance, we verified that the use the exponent 5 instead of 3 in the non linear 
dissipation term gives also similar solutions.
Moreover, as we showed in [11], the introduction of a slowly time variable $J(t)$, 
that requires at least an additional equation, makes possible to obtain solutions 
very similar to other variablity classes.
The system (1) should therefore be considered as one of the simplest tool for
mathematical description of the $\rho$ class bursting, whose properties depends upon
the parameter $J$.
In particular, our aim is to study under which conditions the solutions computed 
for different values of $J$ can be used for describing the transition to stable states, 
like those actually observed in \grs.
 
In particular, the $\chi$ class of Belloni et al. [3] is a typical case in which the 
source X-ray flux remains nearly constant and that, according these authors, it is 
much more frequently observed the all the other ones.
The main purpose of the present paper is to establish conditions for the global 
asymptotic stability of the system (1) and how to calculate the $J$ value corresponding
to the transition from stability to bursting.

\section{Numerical solutions for the bursting behaviour}

Solutions of the system (1) were numerically computed by means of a Runge-Kutta 
fourth order integration routine [14] considering all the four parameters as 
constant, although they, at least in principle, could change in time with the 
physical state of the source.
Our first aim was to obtain a set of parameters' values for which the $x$ variable 
gave a satisfactory solution for the time signals. 
The comparison between the mean burst and energy profiles with the $x$ and $y$ 
curves is shown in the two panels of Fig. 2, where the observed data (left panel) and 
the results of numerical computations (right panel) are shown.
All the curves, after subtraction of their central values, were normalised to 
have close amplitudes. 
The agreement, although not exact, is fully sastifactory considering the rather simple
mathematical model.
The adopted values of the four parameters were 
$\rho = 0.30$, $\chi = 33.0$, $\gamma = 222.0$, $J = 1100.0$.

A very important and unexpected finding of this solution (see [11]), is that the variable $y$
reproduces well the evolution of mean photon energy, that is related to the disk 
temperature.
This finding makes stronger the heuristic meaning of the non-linear model and 
suggests a simple linear relation between the variable $x$ with the source luminosity
$L$, or any other physical quantity proportional to it as the photon density in the
emitting regions of the disk, and of $y$ with the mean energy of photons, otherwise 
the main features of both data series would not be so finely matched.
This correspondence between the formal variables $x$, $y$ and the physical observable 
quantities, is also apparent from the $x,y$ plot (hereafter phase space plot) that 
results similar, except for the scale factors, to the count-rate vs mean energy plot 
studied by Massa et al. [10] (see also Fig. 14 in Janiuk \& Czerny [15]).

Equilibrium points are the intersections of nullclines which are the curves given by
equations for $(dx/dt) = (dy/dt) = 0$, that for Eqs.(1) are:
\begin{eqnarray}
 \gamma y &=& - \rho x^3 + \chi x - J    \nonumber \\
  x &=& y  
\end{eqnarray}
\noindent
and that can be easily reduced to the cubic equation:

\begin{equation}
   \varphi (x) = x^3 + \frac{\gamma - \chi}{\rho}~ x + \frac{J}{\rho}  =  0
\end{equation}

Considering the parameters' values given above there is a unique (negative) 
equilibrium point given by:

{\setlength{\mathindent}{0cm}
\begin{equation}
 x_* = \sqrt[3]{-\frac{J}{2\rho} + \sqrt{\Big(\frac{J}{2\rho}\Big)^2 +\Big(\frac{\gamma - \chi}{3\rho}\Big)^3 }} ~+~ 
       \sqrt[3]{-\frac{J}{2\rho} - \sqrt{\Big(\frac{J}{2\rho}\Big)^2 +\Big(\frac{\gamma - \chi}{3\rho}\Big)^3 }} 
\end{equation}
}

\noindent
that, using the above parameters' values, gives the numerical value $x_* = y_* = -$5.54891.

Note that for about all the $SLT$ the $x, y$ trajectory remains very close to the $x$ 
nullcline, and only after it approaches the other nullcline near the equilibrium point, 
the trajectory rapidly moves to the right to reach the $x$ maximum on the other branch 
of the cubic line and then it decreases very rapidly towards the minimum on the former 
branch.

A transition from bursting to a stable state is obtained in our model simply with a 
change of the $J$ value.
We will show in Sect.4 how the changes of this parameter affect the stability 
of the solution and that these modifications can account for the observed variability 
of \grs.

\section{Stability analysis}

\subsection{Linear analysis}

Consider the system (1) with the assumptions (2),
there is only a unique real solution $x_* < 0$ and therefore only one 
equilibrium point $(x_*, y_*)$, $y_*=x_*$, exists and because $x_*$ is the unique negative 
solution of the equation $\varphi(x) = 0$, in the following we will use $x_*$ as 
parameter instead of $J$.

\begin{figure} [tb]
\centering
\includegraphics[height=7.0cm,angle=-90]{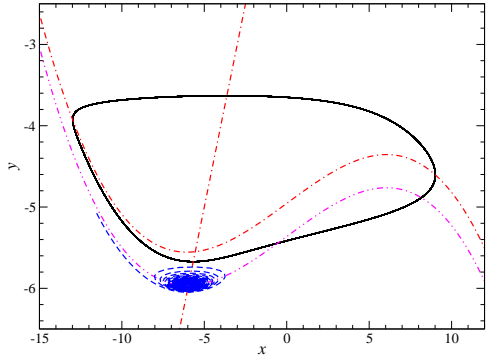}
\caption[]{
Phase space plot of the periodic solution of the system (1) computed using the same
parameters' values of curves Fig. 2. The loop is described in anti-clockwise direction
as observed in the count-rate vs mean energy plots discussed by Massa et al. [10].
Red dash-dot lines are the {\it nullclines} for the Eqs (1).
The violet dashed dot-dot line is the cubic equation for $J$=1190.6 and the dashed blue line is the
corresponding solution of Eqs.(1), showing a trajectory approaching the stable 
equilibrium point. 
}
\label{fig3}
\end{figure}

For the study of the stability it is more convenient to define the two 
new variables:
\begin{eqnarray}
  u &=& x - x_*     \nonumber \\
  v &=& (y - y_*) - (x - x_*)/\gamma 
\end{eqnarray}

The system of differential equations becomes:
\begin{equation}
\left\{
\begin{array}{lcl}
  du/dt &=& \gamma [f(u) -v]   \\ 
  dv/dt &=& g(u)         
\end{array}\right.
\end{equation}
\noindent
where defining $ r = \rho/ \gamma$ ~~ and ~~ $ a = 3 x_*^2 +1/\rho - \chi / \rho$ ~it results
\begin{equation}
f(u) = - r u [u^2 + 3x_* u + a]  
\end{equation}
\noindent
and 
\begin{equation}
g(u) =  r u [u^2 + 3x_* u + a + (1- 1/\gamma)/r] ~~ ~~.
\end{equation}
\noindent
Taking into account the conditions (2) we have
\noindent
\begin{equation}
\gamma > 1, r > 0~~,~~~ x_* < 0~~, ~~~ x_*^2 < (4/9) [a + (1- 1/\gamma)/r]~~~~~~.  
\end{equation}

From the linear analysis [11, 16] one obtains the following results:

\noindent
1) if $a > 0$, $O(0,0)$ is locally asimptotically stable; \\
\noindent
2) if $a < 0$, $O(0,0)$ is unstable.

\noindent
In the plane $(a, x_*^2)$ (see Fig. 4), we ~are inside~ the region~ defined by 
$a > -(1- 1/\gamma)/r$ and $0 < x_*^2 < (4/9) [a + (1- 1/\gamma)/r]$ and we
consider the two regions {\bf A} and {\bf B} defined by the following 
open sets:

\begin{itemize}
\item
{\bf A}: $\{ (a, x_*^2)~ | - (1- 1/\gamma)/r < a < 0 , 0 < x_*^2 < (4/9) [a + (1- 1/\gamma)/r] \}$, 
in which there is a unique unstable equilibrium point;

\item
{\bf B}: $\{ (a, x_*^2)~ | a > 0  , 0 < x_*^2 < (4/9) [a + (1- 1/\gamma)/r] \}$,
in which there is a unique locally asimptotically stable equilibrium point;

\end{itemize}

\noindent
On the segment  $x_*^2 = (4/9) [a + (1- 1/\gamma)/r]$ , $-(1- 1/\gamma)/r < a < 0$ there are two
unstable equilibrium points, while on the half-line 

~

$x_*^2 = (4/9) [a + (1- 1/\gamma)/r]$ , $ a > 0$ 

~

\noindent
there are again two equilibrium points, one of which is unstable and the other one is locally 
asimptotically stable.
The linear analysis, however, does not give information about the stability of the points on the 
segment with $a = 0$. 

\subsection{Lyapunov function analysis}

In [11] we demonstrated that if $(a, x_*^2) \in {\bf A}$, 
there exist at least a limit cycle of (7) around the equilibrium point $O(0, 0)$.
In the following we demonstrate a general theorem to obtain sufficient conditions
for the global asymptotic stability of the system (1).
To this aim we introduce a family of biparametric Lyapunov functions that for this particular
case generalizes the families introduced by Ardito \& Ricciardi [17]. 
The proof is reached following La Salle \& Lefschetz [18].
~~\\
~~\\
{\bf Theorem 1}: ~~{\it  Let us consider the system} (7) {\it where $f(\cdot)$ and $g(\cdot)$ 
are defined in} (8) {\it  and} (9), {\it respectively, under the conditions} (10). 
{\it If $a \geq 0$ and if there exist $\tau, \nu \in \mathbb{R}_+ \cup \{0\}$ such that

\begin{equation}
g(u) \cdot [e^{\nu \int_0^u g(s) ds} (\tau -f(u)) - \tau] \geq 0 
\end{equation}
\noindent
then it follows that $O(0,0)$ is a globally asimptotically stable equilibrium 
point.}
~~~ \\

{\it Proof}.~
Let 
\noindent
\begin{displaymath}
W_{\tau, \nu}: ~~ \mathbb{R}^2 \to \mathbb{R}
\end{displaymath}

\noindent
and

\noindent
\begin{equation}
W_{\tau, \nu}(u,v) = \int_0^v t e^{\nu \gamma t(t/2 - \tau)} d t + \frac{1}{\gamma} e^{\nu \gamma v(v/2 - \tau)} \int_0^u g(t) 
e^{\nu \int_0^t g(s) d s } d t
\end{equation}
\noindent
obviously  $W_{\tau, \nu} \in C^1(\mathbb{R}^2,\mathbb{R})$ and $W_{\tau, \nu}(0,0) = 0 $. 

Being $\gamma > 1$ and  $u g(u) >0$ $\forall u \neq 0$, it follows that 

~

$W_{\tau, \nu}(u, v) > 0 ~\forall~ (u, v) \in \mathbb{R}^2 -\{O(0,0)\}$ ~~~~.

~

Moreover from the hypothesis (11) we get
\begin{equation}
\dot{W}_{\tau, \nu}(u,v) = - e^{\nu \gamma v(v/2 - \tau)} g(u) [(\tau - f(u)) e^{ \nu \int_0^u g(s) d s } - \tau] \leq 0 ~~.
\end{equation}

\noindent
where $\dot{W}_{\tau, \nu}$ is the derivative of $W_{\tau, \nu}$ along the trajectories of the system (7);
therefore  $W_{\tau, \nu}$ is a Lyapunov function and $O(0,0)$ is stable.

In order to prove that $O(0,0)$ is globally asymptotically stable we observe that $W_{\tau, \nu}$
is {\it radially unbounded}, given that from (10)
{\setlength{\mathindent}{0cm}
\begin{flalign}
\lim_{\parallel (u,v) \parallel \rightarrow +\infty} W_{\tau, \nu}(u,v) = +\infty ~,
\nonumber
\end{flalign}}

\noindent
thus the set
\begin{equation}
S_l = \{(u,v) \in \mathbb{R}^2 | W_{\tau, \nu}(u,v) < l, l \in \mathbb{R}_+ \}
\nonumber
\end{equation}
\noindent
is bounded.

Let
\begin{equation}
E = \{(u,v) \in S_l | \dot{W}_{\tau, \nu}(u,v) = 0 \}
\nonumber
\end{equation}
\noindent
being $\{O(0,0)\}$ the largest invariant set in $E$ the claim follows.

On the basis of this theorem it is possible to establish the following conditions for
the global asymptotical stability.

~~\\
{\bf Corollary 1}:~~{\it  Let us consider the system} (7) {\it where $f(\cdot)$ and $g(\cdot)$ are defined} (8) 
{\it and} (9),{\it respectively, under the conditions} (10).
{\it 
If 

$a \geq 0$,~~~ and ~~~ $ x_*^2 < (4/9)a $ 

\noindent
then $O(0,0)$ is globally asimptotically stable.}

{\it Proof}.~
In this case one has $u \cdot f(u) \leq 0$ and the thesis is a direct consequence of Th. 1
by taking $\tau = \nu = 0$ 
in (12) and observing that the condition (11) is satisfied.
~~\\ 

Note that the family of Lyapunov functions given in (12), with the particular choice
$\tau = \nu = 0$ was introduced by Hsu [19].
This particular function, however, does not work in the analysis of the case for $a = 0$
that corresponds to the transition between the regions of stablity and instability 
of the equilibrium point.
For the study of this case we must consider again the Lyapunov function (12).
Thus we have:
~~\\

\noindent
{\bf Corollary 2}:~~{\it  Let us consider the system} (7) {\it where $f(\cdot)$ and $g(\cdot)$ are defined 
in} (8) {\it and} (9), {\it respectively, with the conditions given in} (10).
{\it 
Let

~

$b \doteq  a + (1 - 1/\gamma)/r $, 

\noindent
if 

$ a \geq 0 $ ~~and~~ $ \frac{4}{9} a < x_*^2 < \frac{b}{6} + \frac{ab}{9(a+b)} + \frac{1}{9} \sqrt{2ab} $

~~\\
\noindent
then $O(0,0)$ is globally asimptotically stable.}

{\it Proof}.~
We consider the function $W_{\tau, \nu}(u,v)$ defined by (12), then defining

\begin{equation}
S(u) \doteq e^{\nu \int_0^u g(s) d s } [\tau - f(u)] - \tau~~~~
\end{equation}

\noindent
we have that $\forall \tau , \nu > 0$ it follows

\begin{equation}
\lim_{u \rightarrow -\infty} S(u) = -\infty ~~, ~~ \lim_{u \rightarrow +\infty} S(u) = +\infty ~~, ~~ S(0) = 0
\end{equation}

\noindent
and

\begin{equation}
\frac{dS(u)}{du} = e^{\nu \int_0^u g(s) d s } ~T(u) ~~~~,
\end{equation}

\noindent
where

\begin{equation}
T(u) = \nu g(u) (\tau -f(u)) - f'(u)~~~~.
\end{equation}

With the choices $\tau = -3 x_* (a+b) r$ and $\nu = 2/rb(a+b)$, from the defintions
of $f(u)$ and $g(u)$, one has

{\setlength{\mathindent}{0cm}
\begin{eqnarray}
T(u)  & = & \frac{2r}{b(a+b)} \Big\{ u^4 (u + 3x_*)^2 + (a+b) [u^4 + 9 u^2 ((ab/9)(a+b) +  \nonumber \\ 
         &  & + b/6 -x_*^2) + ab/2] \Big\}   
\end{eqnarray}}

From our assumptions it follows $T(u) \geq 0$, the condition (11) is satisfied and consequently 
the thesis is proved.

\begin{figure} 
\centering
\includegraphics[height=6.3cm,angle=0,scale=1.0]{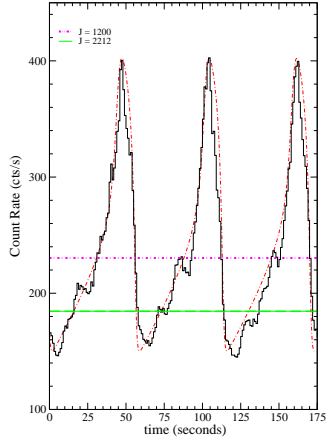}
\includegraphics[height=6.3cm,angle=-90,scale=1.0]{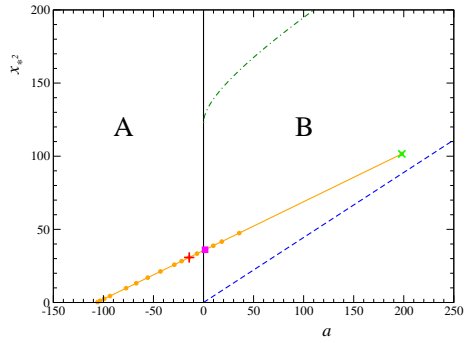}
\caption[]{
(Left panel: A short segment of the data of the same observation of those in the upper panel of Fig. 1 
superposed to the solution of our mathematical model with $J = 1100$ (red dash-dot curve). 
The two horizontal lines correspond to the signal levels calculated with $J = 1200$ (magenta dotted
line) and $J = 2212$  to match the data of the observation of November 1996 when the source 
was in a stable state whose mean value is represented by the thick dashed green line. 
Right panel: Detail of the region in the $(a, x_*^2)$ plane where is the solution for data modelling 
described in the text.
The dashed blue line is given by the condition $x_*^2 = (4/9)a$ and the green dash dotted curve corresponds to
the upper limit for $x_*^2$.
The orange continous line gives the locus of equilibrium points for different values of $J$, unstable in the region 
A and stable in region B, the red $\times$ marks the point corresponding to the red curve in the left
panel, the magenta $\times$ corresponds to $J = 1200$ and the green $\times$ gives the level  
observed in November 1996.
}
\label{fig4}
\end{figure}

\section{The transition from stable to unstable equilibrium}
 
In addition to the description of the burst profiles of \grs, the system (1)  can be 
also used to study the transition to stable classes.
Following Massaro et al. [11], we consider that $\rho$, $\chi$ and $\gamma$  
remanin constant while the transition is driven by a slow change of $J$, that one can expect 
to be related with the accretion mass rate of the disk and then to its emission power.

The left panel of Fig. 4 shows the computed burst profile using the parameters' values given
in Sect. 2 together with a data segment in the energy range [1.6, 10] keV and
the thick dashed green line indicates the mean level measured when the source was 
in the stable $\chi$ class observed by \sax.
Let us observe that from the definition of $ a = 3 x_*^2 +1/\rho - \chi / \rho$ and from eq. (4)
it results that
\begin{equation}
J = -x_* [ a + (1 - \frac{1}{\gamma})\frac{1}{r} - 2 x_*^2 ] ~~~~.
\nonumber
\end{equation}

\noindent
The right panel shows a region in the plane $(a, x_*^2)$ in which it is located the 
point corresponding to the model and the orange line tracks the locus of equilibrium points 
when $J$ increases from 100 (left lower extreme) to 2212 (right upper extreme).
Notice that this line crosses the axis $a = 0$ for the critical $J_c = 1190.6$ at the 
transition point between unstable and stable equilibrium, in which the limit cycle 
converges soon to a constant value.
This effect is illustrated in Fig. 3 by the blue lines: the dashed line is the nullcline 
for this critical value and the trajectory, computed using Eqs. (1), describes the 
global asymptotic stability of this point, as demonstrated in the previous section.
Note that the orange line is entirely inside the region limited by the the dashed lines
that are the limits for the application of the Cor. 2.
Then all the solutions for $J > J_c$ are globally asymptotically stable and describe the
$\chi$ class with a decreasing flux level for increasing $J$.

\begin{figure} 
\centering
\includegraphics[height=8.5cm,angle=0,scale=1.0]{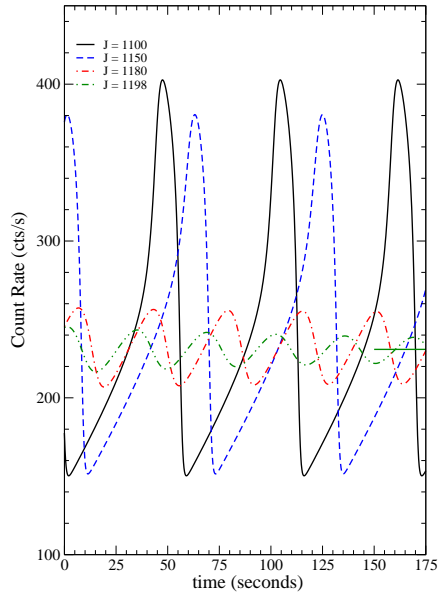}
\caption[]{
Evolution of the signal shape in the transition to stability: the black continuous curve is the one used to model the data and the other curves have been calculated for different values of $J$ on the two sides of the transition line.
Note the changes of the amplitude and of the recurrence time approaching $J_c$, and for $J > J_c$ the progressive decrease of the amplitude towards the stable value, represented by the continuous  horizontal segment on the right side.   
}
\label{fig5}
\end{figure}

We can investigate the changes of the signal for $J$ varying across the transition  to stability. 
Some examples of burst pattern computed by means of our model are given in Fig. 5 in 
which we applied the same scale factors used for the left panel of  Fig. 4.
One can see that the amplitude of the signal has a fast decrease for $J$ just lower than the
critical value, the burst profile is more symmetric without the sharp increase and the 
recurrence time becomes shorter.
For $J$  just above $J_c$, a rather fast decay of the burst amplitude to the stable level, 
represented by the horizontal segment in Fig. 5, is apparent; higher values of $J$ correspond
to faster decaying to the equilibrium.
In these cases, however, because of the fast reaching of equilibrium, the resulting evolution 
of the signal is largely dependent on the initial values of the variables, and particularly, 
how much they differ form the equilibrium values.
\\
A similar behaviour is obtained also for $y$, but considering the rather low amplitude of
the observed modulation (about 10\% in the lower panel of Fig. 1), a decrease to a few percent
would require a very high instrumental sensitivity for a statistically significant detection.

The comparison of the class transition derived from the model with the observed behaviour of 
\grss is practically very hard because it requires the detection of the transition from the $\rho$ to the $\chi$ class.
This transition depends on the rapidity of the $J$ variation and could occur in a rather short
time, then the probability that an X-ray telescope on board an orbiting satellite is pointing at 
the source just in such an occasion is very low.
In particular, no $\rho-\chi$ transition is observed in the entire \sax~ data set considered in our 
analysis.

\section{Conclusion}

Non-linear dynamical models can be used for reproducing several interesting features of the time 
behaviour exhibited by the variable astrophysical source \grs.
In this paper we focused our interest on the study of stability of the solutions of the system proposed
by Massaro et al. [11] for the $\rho$ class bursting and explored the transition between this and another
class characterized by a stable emission.
To this aim we introduced a biparametric Lyapunov function, that extends the previous results
of Ardito \& Ricciardi [17].
We found that, in the more physically interesting region of the parameters' space, it is possible to 
have either an unstable equilibrium point, around which closed trajectories of limit cycle can be 
established, or equilibrium points, whose global asymptotical stable was demostrated in Corollary 2.
The observed bursting behaviour of the source and the transition to stable states and vice versa, can 
be thus simply accounted for by changes of only the forcing parameter.

In the past years two other sources have been observed to exhibit a similar bursting 
pattern: IGR~J17091+3624 (Altamirano et al. [20]), and the X-ray binary MXB 1730-335 (also known as Rapid 
Burster) very recently reported by Bagnoli \& in't Zand [21], the latter one with a long recurrence 
time of about 7 minutes, suggesting that it could be a phenomenon more frequent than considered in the 
past when \grss was the unique cosmic source.
The approach presented here can be useful for developing general mathematical tools for  
modelling  this complex quasi-periodic process in astrophysical sources and to obtain a description
in terms of observable quantities of the accretion disk instabilities.

~~~~ \\

{\bf References}

\clearpage

\end{document}